\pdfoutput=1
\documentclass[12pt,letterpaper,floatfix]{article}
\usepackage{amsmath}
\usepackage{amssymb}
\usepackage{epsfig}
\usepackage{epstopdf}
\usepackage{euscript}
\usepackage{setspace}
\usepackage{float}
\usepackage[export]{adjustbox}
\usepackage{color}
\definecolor{splashwhite}{rgb}{1.0,0.99,1.0}

\def\mathswitchr#1{\relax\ifmmode{\mathrm{#1}}\else$\mathrm{#1}$\fi}

\def\Order#1{${\cal O}(#1)$}
\def\Ordpr#1{${\cal O}(#1)_{prag}$}

\def\rQCED{{\rm QCED}}

\newcommand{\KK}{${\cal KK}$}

\newcommand{\Meu}{\EuScript{M}}

\def\st{\hbox{}} 
\newcommand {\KKMC}{\hbox{${\cal KK}$}\ MC}

\newcommand {\pslash}{\hbox{$\not\hbox{\kern-2.3pt $p$}$}}

\usepackage{epsfig}

\usepackage{cite}

\usepackage{epic}

\def\Order#1{${\cal O}(#1$)}
\def\Ordpr#1{${\cal O}(#1)_{prag}$}

\def\alf1{ {\alpha\over\pi} }
\def\rQCED{{\rm QCED}}
\begin{document}
\begin{titlepage}
\begin{center}
{\bf \large {\KK}MC-hh: Resummed Exact ${\cal O}(\alpha^2L)$ EW Corrections in a Hadronic MC Event Generator}\\
\vspace{2mm}
  {\bf   S. Jadach\\
Institute of Nuclear Physics Polish Academy of Sciences, PL-31342 Krakow, Poland\\
   E-mail: stanislaw.jadach@ifj.edu.pl\\}
{\bf B.F.L. Ward\\%
      Baylor University, Waco, TX 76798, USA\\
        E-mail: bfl\_ward@baylor.edu\\}
{\bf Z.A. Was\footnote{The work is supported in part by 
the  Programme  of the French–Polish  Co-operation between IN2P3 and COPIN within the Collaborations No.10-138.}
\\
 Institute of Nuclear Physics Polish Academy of Sciences, PL-31342 Krakow, Poland\\
 E-mail: zbigniew.was@ifj.edu.pl\\}

{\bf S.A. Yost\footnote{Work supported in part by 
grants from The Citadel Foundation.}\\
      Institute of Nuclear Physics Polish Academy of Sciences, PL-31342 Krakow, Poland and 
      The Citadel, Charleston, SC, USA\\
        E-mail: scott.yost@citadel.edu\\}
\end{center}
\vspace{2mm}
\centerline{\bf Abstract}
We present an improvement of the MC event generator Herwiri2, where we recall the latter MC was a prototype for the inclusion of CEEX resummed EW corrections in hadron-hadron scattering at high cms energies. In this improvement the new exact ${\cal O}(\alpha^2L)$ resummed EW generator \KK MC 4.22, featuring as it does the CEEX realization of resummation in the EW sector, is put in union with the Herwig parton shower environment. The {\rm LHE} format of the attendant output event file means that all other conventional parton shower environments are available to the would-be user of the resulting new MC. For this reason (and others -- see the text) 
we henceforth refer to the new improvement
of the Herwiri2 MC as {\KK}MC-hh. Since this new MC features exact ${\cal O}(\alpha)$ pure weak corrections from the DIZET EW library and features the CEEX and the EEX YFS-style resummation of large multiple photon effects, it provides already the concrete path to 0.05\% precision on such effects if we focus on the EW effects themselves. We therefore show predictions for observable distributions and comparisons with other approaches in the literature. This MC represents an important step in the realization of the exact amplitude-based $QED\otimes QCD$ resummation paradigm. Independently of this latter observation, the MC rigorously quantifies important EW effects in the current LHC experiments.
\vspace{1cm}
\begin{center}
 BU-HEPP-16-02,  IFJ-PAN-IV-2016-20, June, 2016\\
\end{center}

\end{titlepage}
%

 
\def\Kmax{K_{\rm max}}\def\ieps{{i\epsilon}}\def\rQCD{{\rm QCD}}

\section{\bf Introduction}\par

In the current era of precision QCD, 
by which we mean
predictions for QCD processes at the total precision tag of $1\%$ or better,
it is important to have rigorous baselines with respect to which to compare 
theoretical results both for the QCD theory and for the corresponding EW theory
which needs to be calculated to the respective attendant order of precision
when a specified level of precision is envisioned for the QCD corrections. In this context,
we have developed the theoretical paradigm based on the exact amplitude-based resummation of both large EW effects and large QCD effects, $QED\otimes QCD$ resummation as presented in Refs.~\cite{qced1,qced2,qced3,qced4,qced5,qced6,qced7,qced8,qced9}. The first publicly released MC featuring the resummation of large
QCD effects in this framework was Herwiri1.031~\cite{herwiria,herwirib,herwiric,herwirid,herwirie,herwirif,herwirig}, an IR-improved version of Herwig6.5~\cite{hwg65-1,hwg65-2}, and it was shown to give a better description of the
$Z/\gamma^*\; p_T$ spectrum than the correspondingly unimproved Herwig6.5 in the regime below 20GeV where soft effects are important.
The first MC featuring resummation of large  EW effects in this paradigm was presented in Ref.~\cite{hwri2.0} as Herwiri2.0 and was based on the union the QCD parton shower MC Herwig6.5 and the Monte Carlo \KKMC 4.19~\cite{kkmc419a,kkmc419b}
for large EW effects. \KKMC 4.19 realizes, on an event-by-event basis, CEEX/EEX~\cite{kkmc419a} YFS-style~\cite{yfs-1,yfs-2} resummation of large EW effects to all orders in $\alpha$. In what follows, we improve the development in Herwiri2 by realizing the union of the MC Herwig6.5 
and the new version of the Monte Carlo \KKMC, version 4.22~\cite{kkmc422}, in which the incoming beams can be quarks and anti-quarks, unlike the case in \KKMC 4.19 wherein the incoming beams are set to $e^+$ and $e^-$. We also stress that, as we use \KKMC 4.22 to drive the union and Herwig6.5 to shower
the respective `hard' events, use of the Les Houches Event File format~\cite{lhe-format} for the hard event output means that any QCD shower MC that accepts that format for input hard events can be substituted for Herwig6.5 in realizing the corresponding parton shower effects. To make this more evident, we henceforth refer to the attendant improvement
of Herwiri2 as {\KK}MC-hh, where `hh' denotes that it simulates processes involving incoming hadron-hadron beams\footnote{In Ref.~\cite{say-llgs16}, {\KK}MC-hh was denoted as Herwiri2.1.}. We will the denote
by {\KK}MC-ee the new version 4.22 of the {\KK}MC which features incoming $e^+e^-$ beams with beamsstrahlung as needed for precision studies  for
$e^+e^-$ colliding beam devices. \par

Continuing with the motivation and perspective, we recall in Refs.~\cite{EWsys1,EWsys2,EWsys3,Horace1,Horace2,Horace3} that it has been shown that EW corrections,
depending of course in detail on the respective cuts, can be easily at the several per cent level in the ATLAS and CMS and LHCb 
acceptances for the production of single $Z/\gamma^*$'s with the decays to lepton pairs. We note that the current statistical accuracy in the 7TeV and 8TeV data samples are, for 7TeV, 0.06\%, 0.06\% and 0.45\%  and, for 8TeV, 0.03\%, 0.03\% and 0.3\%, respectively, and that the total error, excluding the luminosity contribution, on these processes in distributions such as the $Z/\gamma^*$ $ p_T$ spectrum is at the 0.5\% level~\cite{atlaspt,cmsymll,lhcbpt} in the low $p_T$ regime with estimates of the EW correction contribution to this error at the $\sim .2\%$  level. The latter estimates are based on the comparisons of available exact ${\cal O}(\alpha)$ EW corrections 
with resummed FSR~\cite{Horace1,Horace2,Horace3,photos1,photos2,photos3,photos4,photos5,shpa,hwg++,pythia6}. The theoretical precision required
on the simulation of all EW corrections in these data is at the 0.05\% level -- with such precision, one can cross check the currently used estimates for the size of the EW FSR effects in the LHC experiments and one can check the sizes of the ISR and IFI 
initial state - final state interference effects that may enter as well at the per mille level, depending on the cuts. The event generator \KKMC 4.22, providing as it does the respective exact ${\cal O}(\alpha^2 L)$ CEEX/EEX multiple photon radiative effects, renders such precision for EW corrections for the hard processes at hand. Moreover, it affords a realistic event-by-event simulation of the actual multiple photon radiative effects in the data using exact, amplitude-based resummation so that detector cut effects can be more faithfully simulated accordingly. {\KK}MC-hh, the union of KK MC4.22 and Herwig6.5, is thus the only hadron MC event generator for LHC phyiscs which features the exact ${\cal O}(\alpha^2 L)$ EW correction with CEEX/EEX YFS-style resummation
for ISR, IFI and FSR multiple photon radiative effects at the level of the amplitude. This sets it apart from the standpoint of its theoretical precision tags for EW effects. {\KK}MC-hh is an important step in the realization of amplitude-based
exact ${\cal O}(\alpha_s^2,\alpha_s\alpha,\alpha^2L)$ $QED\otimes QCD$ resummation in a MC event generator.\par

The discussion proceeds as follows. In the next Section, we give a brief review of the relevant aspects of the theory underlying the new MC. In Section 3, we show some sample MC data and comparisons with other approaches in the literature. Section 4 sums up; it also describes how to obtain the MC. Detailed comparisons with LHC data will appear elsewhere~\cite{elswh}.\par

\section{Brief Review of the Theoretical Foundations of \\
              {\KK}MC-hh}

In this section we give a brief review of the theory that underlies {\KK}MC-hh. The key ingredient is the new version of the \KKMC, version 4.22. It allows the incoming beams to be $f\bar{f}, \; f=q,\;\ell,\; q=u,d,s,c,b, t, \ell=e,\mu,\tau, \nu_e,\nu_\mu,\nu_\tau$.
Given that the  CEEX/EXX~\cite{kkmc419a,kkmc419b,kkmc422} realizations of YFS exponentiation are used, let us briefly recall the corresponding theory, as it remains one that is not generally known.\par

For the prototypical hard process, $q\bar{q}\rightarrow \ell\bar{\ell}+n\gamma, \; q=u,d,s,c,b,t,\ell=e,\mu,\tau,\nu_e,\nu_\mu,\nu_\tau,$ we have the master formula
\begin{equation}
\sigma =\frac{1}{\text{flux}}\sum_{n=0}^{\infty}\int d\text{LIPS} \rho_A^{(n)}(\{p\},\{k\}),
\label{eqn-hw2.1-1}
\end{equation}
where $A=\text{CEEX},\;\text{EEX}$, where we use the abbreviated notation $\{p\}$ for the incomong and outgoing fermion momenta and  $\{k\}$ for the $n$ photon
four momenta, and where we have, from Refs.~\cite{kkmc419a,kkmc419b,kkmc422}, for example,
\begin{equation}
\rho_{\text{CEEX}}^{(n)}(\{p\},\{k\})=\frac{1}{n!}e^{Y(\Omega;\{p\})}\bar{\Theta}(\Omega)\frac{1}{4}\sum_{\text{helicities}\;{\{\lambda\},\{\mu\}}}
\left|\Meu\left(\st^{\{p\}}_{\{\lambda\}}\st^{\{k\}}_{\{\mu\}}\right)\right|^2
\label{eqn-hw2.1-2}
\end{equation}
with an analogous formula for the case $A=\text{EEX}$ as it is given in Refs.~\cite{kkmc419a,kkmc419b,kkmc422}. Here $\text{LIPS}$ denotes Lorentz-invariant phase-space. The YFS infrared exponent $Y(\Omega;\{p\})$, the attendant infrared integration limits specified by the region $\Omega$ and its characteristic function
$\Theta(\Omega,k)$ for a photon of energy $k$, with $\bar\Theta(\Omega;k)=1-\Theta(\Omega,k)$ and $$\bar\Theta(\Omega)=\prod_{i=1}^{n}\bar\Theta(\Omega,k_i),$$ as well as the CEEX amplitudes $\{\Meu\}$ are all given in Refs.~\cite{kkmc419a,kkmc419b,kkmc422}.\par

In the {\KK}MC 4.22, the exact EW corrections are implemented using the DIZET6.2.1 EW library from the semi-analytical
program ZFITTER~\cite{zfitter1,zfitter2}. The implementation steps are described in Ref.~\cite{kkmc419a} so that we do not repeat them here.\par

The union with the parton shower in Herwig6.5~\cite{hwg65-1,hwg65-2} proceeds via the standard formula for the Drell-Yan process:
\begin{equation}
\sigma_{\text{DY}}=\int dx_1dx_2\sum_i f_i(x_1)f_{\bar{i}}(x_2)\sigma_{\text{DY},i\bar{i}}(Q^2)\delta(Q^2-x_1x_2s),
\label{eqn-hw2.1-3}
\end{equation}
where the subprocess for the $i$-th $q\bar{q}$ annihilation with $\hat{s}=Q^2$ when the pp cms energy squared is $s$
is given in a conventional notation for parton densities $\{f_j\}$. The backward evolution~\cite{sjos-sh} 
for the densities as specified in
(\ref{eqn-hw2.1-3}) then gives {\KK}MC-hh multple gluon radiation and the attendant hadronization for the that shower.
We use in what follows the Herwig6.5 shower MC for this phase of the event generation. We stress however that, as the Les Houches Accord format is also available for the hard processes generated in {\KK}MC-hh before the shower, all shower MC's which use that format can be used for the shower/hadronization part of the simulation. Studies with such other choices will appear elsewhere~\cite{elswh}.\par

The event generation itself proceeds as follows. The adaptive MC FOAM~\cite{sj-foam1,sj-foam2} calculates the primary distribution of quarks 
and ISR photons to set up an appropriate distribution grid during an exploratory phase at the beginning of the run. A four dimensional distribution generates the quark flavor, the hard process scale $Q$, one of the light-cone fractions $x_i$, and the amount of ISR photon radiation in a convenient unit of measure. These generations are mapped into the generation of four random numbers in the interval $[0,1]$. The first is uniformly distributed between u, d, c, s and b quarks and anti-quarks flavor indices. The remaining three are in a 3-dimensional volume which
is mapped into simplicial cells to optimize the MC integration. There is no need for sophisticated mapping before calling FOAM, though
some minimal mapping is done, since an exponential map for $x_i$
improves performance.\par

In closing this section, we note that, with the generation of multiple gluon effects via Herwig6.5 as described above for the hard events generated by the {\KK}MC modules in {\KK}MC-hh, we realize exactly the terms ${\cal O}(\alpha_s^nL^n, \alpha^2L')$
and that part of the terms  ${\cal O}(\alpha_s^nL^n\alpha^2L')$ which factorizes. Here, $n= 0,1,2,\ldots$ and $L,L'$ are the respective QCD and QED big logs. The results in Ref.~\cite{dittmaier-1,dittmaier-2} show
that at ${\cal O}(\alpha_s\alpha)$ the nonfactorizable part is a small one in general. Let us recall the master formula for 
$QED\otimes QCD$ resummation~\cite{qced1,qced2,qced3,qced4,qced5,qced6,qced7,qced8,qced9,irdglap1,irdglap2}:\newpage
\begin{eqnarray}
&d\bar\sigma_{\rm res} = e^{\rm SUM_{IR}(QCED)}
   \sum_{{n,m}=0}^\infty\frac{1}{n!m!}\int\prod_{j_1=1}^n\frac{d^3k_{j_1}}{k_{j_1}} \cr
&\prod_{j_2=1}^m\frac{d^3{k'}_{j_2}}{{k'}_{j_2}}
\int\frac{d^4y}{(2\pi)^4}e^{iy\cdot(p_1+q_1-p_2-q_2-\sum k_{j_1}-\sum {k'}_{j_2})+
D_\rQCED} \cr
&\tilde{\bar\beta}_{n,m}(k_1,\ldots,k_n;k'_1,\ldots,k'_m)\frac{d^3p_2}{p_2^{\,0}}\frac{d^3q_2}{q_2^{\,0}},
\label{subp15b}
\end{eqnarray}\noindent
where $d\bar\sigma_{\rm res}$ is either the reduced cross section
$d\hat\sigma_{\rm res}$ or the differential rate associated to a
DGLAP-CS~\cite{dglap1,dglap2,dglap3,dglap4,dglap5,dglap6,cs1,cs2,cs3,cs4} kernel involved in the evolution of PDF's and 
where the {\em new} (YFS-style~\cite{kkmc419a,kkmc419b,yfs-1,yfs-2,kkmc422,yfs-jw1,yfs-jw2,yfs-jw3,yfs-jw4,yfs-jw5,yfs-jw6,yfs-jw9,yfs-jw10,yfs-jw11,yfs-jw12,yfs-jw13}) {\em non-Abelian} residuals 
$\tilde{\bar\beta}_{n,m}(k_1,\ldots,k_n;k'_1,\ldots,k'_m)$ have $n$ hard gluons and $m$ hard photons and we show the final state with two hard final
partons with momenta $p_2,\; q_2$ specified for a generic $2f$ final state for
definiteness. See Refs.~\cite{qced1,qced2,qced3,qced4,qced5,qced6,qced7,qced8,qced9,irdglap1,irdglap2} for the precise definitions of the infrared functions $SUM_{IR}(QCED),\; D_\rQCED$ and for precise definitions of the residuals. In the language encoded in (\ref{subp15b}), {\KK}MC-hh now realizes the EW contributions to the residuals
with exponentiated ${\cal O}(\alpha^2L')$ accuracy and the QCD contributions  to the  residuals to leading log accuracy to all orders in $\alpha_s$ and thereby obtains the attendant mixed corrections as approximated by their factorized forms. We anticipate adding the QCD exact NLO correction
following the methods in Refs.~\cite{jadach1,jadach2,mcatnlo-1,mcatnlo-2} elsewhere~\cite{elswh}.\par

We turn now in the next Section to sample MC data and comparisons with other approaches in the literature.\par

\section{Sample MC data and Theoretical Comparisons}

In this Section we present sample Monte Carlo data to show the sizes of various EW corrections and we compare with results
obtained from other approaches to these EW corrections that are available in the literature. We start with sample Monte Carlo data.\par
\subsection{Sample Monte Carlo Data}
{\KK}MC-hh can be run without electroweak corrections and photons, in
which case it simply replaces the Herwig6.5 hard process generation mechanism,
without essentially changing the physics. The EW corrections can be added
incrementally to test their effect. We made
test runs with $10^6$ events and test runs of $25\times 10^6$ events, with MSTW2008 PDFs, and with a generator cut $50 \text{GeV} < M_{q\bar{q}} < 200 \text{GeV}$
for a pp cms energy of 8 TeV. Using the Herwig6.521 showers, we get the results for the cross sections which are shown in Tab.~\ref{tab-1} from the $25\times 10^6$ event samples.
\begin{table}[ht]
\caption{Showered tests with Herwig6.521.}
\begin{center}
\begin{tabular}{|c|c|c|c|}
\hline
\text{MC}& \text{EW-CORR}& \text{XSECT}& $\Delta(\text{Rel})$\\
\text{HERWIG6.5}& \text{No Photons} &1039.6$\pm$ 0.2 pb& $\cdots$ \\
\text{{\KK}MC-hh}& \text{No Photons}& 1038.69 $\pm$ 0.08 pb& (-0.09\%)\\
\text{{\KK}MC-hh}& \text{CEEX FSR+EWK}& 986.05 $\pm$ 0.11 pb& (-5.2\%)\\
\text{{\KK}MC-hh}&\text{ CEEX ISR+FSR+EWK}& 986.21 $\pm$ 0.26 pb& (-5.1\%)\\
\text{{\KK}MC-hh}&\text{EEX ISR+FSR+EWK}&985.82$\pm$ 0.26 pb&(-5.2\%).\\
\hline
\end{tabular}
\end{center}
\label{tab-1}
\end{table}
Here, we use an obvious notation: $\text{MC}$ denotes which of Herwig6.5210 and {\KK}MC-hh is being used, $\text{EW-CORR}$ denotes the type of EW correction option chosen, XSECT denotes the corresponding cross section result, and $\Delta(\text{Rel})$ denotes the percentage change relative to the reference cross section which is taken here to be that for Herwig6.5210. We denote by the $\text{EW-CORR}$ switch value
$\text{``No Photons"}$ the EW Born level results so that they should agree for {\KK}MC-hh and Herwig6.521 and we see that they do.
The switch value $\text{``FSR+EWK"}$ denotes that we have turned on the exact electroweak \Order{\alpha} corrections in DIZET and the FSR multiple photon radiation from \KKMC~
and the value $\text{``ISR+FSR+EWK"}$ denotes that we have turned on all of the multiple photon radiation from the \KKMC, both ISR and FSR, in addition to the exact \Order{\alpha} electroweak corrections in DIZET. EEX and CEEX denote the respective mode of YFS exponentiation as already noted. We see that the total change of $- 5.2\%$ relative to the reference Born cross section for the $\text{CEEX FSR+EWK}$ result is consistent with what was found already in Refs.~\cite{EWsys1,EWsys2,EWsys3,Horace1,Horace2,Horace3}. We see that the ISR, for this very inclusive selection for $M_{q\bar{q}}$, is a small effect, within the errors of the simulations presented here.
In general, we see from Table~\ref{tab-1} that EW corrections must be calculated accurately in precision LHC physics.
We see that the best EEX and the best CEEX results, with ISR+FSR+EWK corrections,  agree to 0.04\%. Here, if we use the language of Ref.~\cite{kkmc419a},
the best EEX result has exact \Ordpr{\alpha^3} EEX YFS resummation in which the EW  $\tilde{\bar\beta}_{0,m}$ residuals have exact \Order{\alpha^3{L'}^3,\alpha^2 {L'}^2,\alpha^2 L',\alpha L',\alpha} corrections and the best CEEX result has the exact \Order{\alpha^2 L'} correction to exact \Ordpr{\alpha^2} CEEX YFS resummation
in which the analogous residuals have exact \Order{\alpha^2 {L'}^2,\alpha^2 L',\alpha L',\alpha} corrections with IFI. We will feature both in what follows
given their closeness to one another.\par

\begin{figure}[h]
\begin{center}
\setlength{\unitlength}{1in}
\begin{picture}(6,2.4)(0,0)
\put(0,0.2){\includegraphics[width=3in]{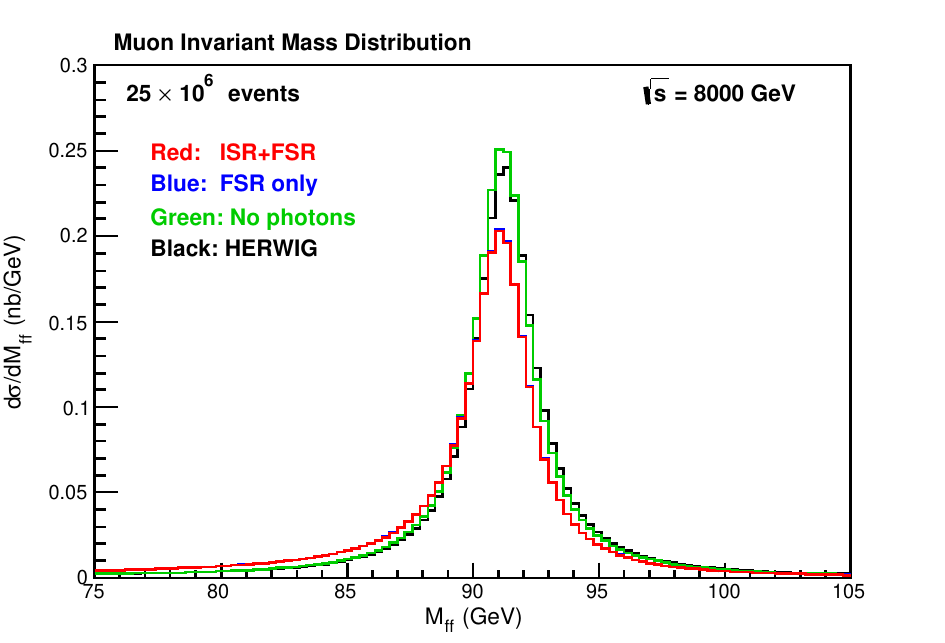}}
\put(3,0.2){\includegraphics[width=3in]{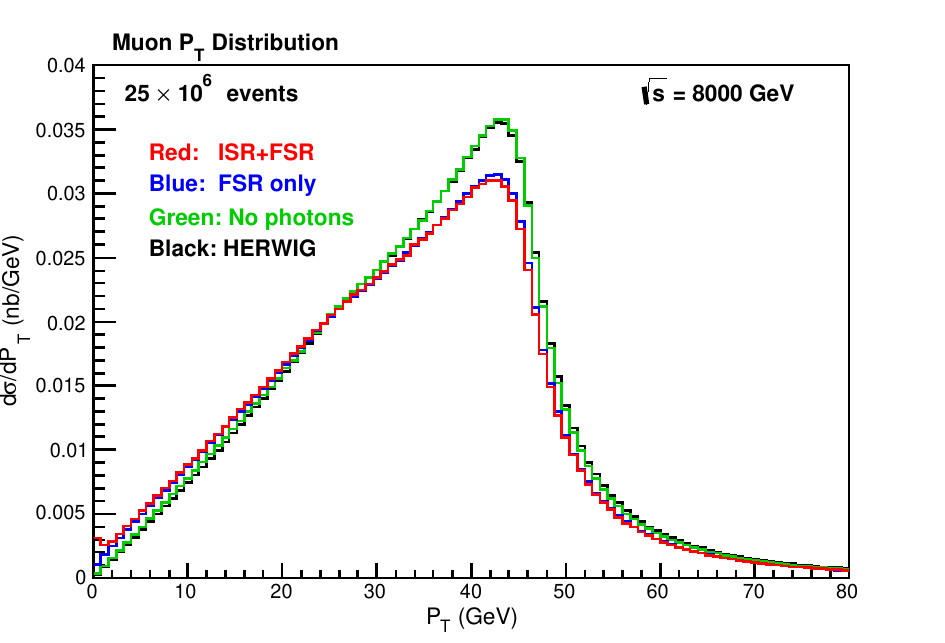}}
\end{picture}
\end{center}
\vspace{-10mm}
\caption{\baselineskip=11pt Muon invariant mass and transverse momentum distributions for {\KK}MC-hh with the cuts specified in the text for the EW-CORR switches ``CEEX ISR + FSR+ EWK" (red -- medium dark shade) $\equiv$ ISR+FSR and ``CEEX FSR+EWR" (blue -- dark shade)$\equiv$ FSR only, showered by HERWIG6.5. The green (light shade) distributions are generated by {\KK}MC-hh with the No photons EW-CORR switch using the HERWIG6.5 hard cross section and the black distributions are made by HERWIG6.5 alone. The switches are explained in the text.}
\label{fig-1}
\end{figure}

We turn next to the muon invariant mass spectrum and the muon transverse momentum ($p_T$) spectrum shown in Fig.~\ref{fig-1}. 
The simulations contain $25\times 10^6$ events and we show results from {\KK}MC-hh for all
three of the $\text{EW-CORR}$ switch values described above, which are denoted in the figure as ``No Photons", ``FSR only" for ``CEEX FSR+EWK" and 
``ISR+FSR" for  ``CEEX ISR+FSR+EWK". In the muon pair mass spectrum, we see the well-known effect of the FSR and ISR to move events from the region above the peak to the region below the peak, where the most pronounced changes are near the peak itself. In the muon $p_T$ spectrum, we see that the black Herwig6.5 curve agrees, as it should, with
the green (light shade) curve for "No Photons" and we see again that
the modulation of the green(light shade) curve for the ``No Photons" case by the ``FSR" and ``ISR+FSR" cases is significant. The effects in Fig.~\ref{fig-1} must be taken into account with accuracy in precision LHC physics~\cite{EWsys1,EWsys2,EWsys3,Horace1,Horace2,Horace3}.\par
   
\begin{figure}[ht]
\begin{center}
\includegraphics[width=160mm]{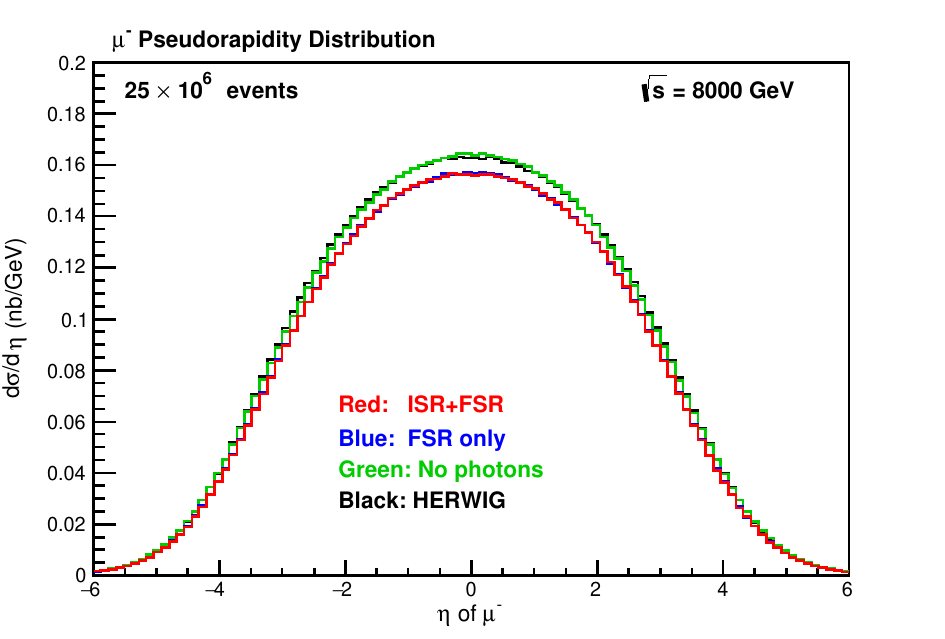}
\end{center}
\vspace{-4mm}
\caption{\baselineskip=11pt The muon $\eta$ spectrum in {\KK}MC-hh with the cuts specified in the text. The green(light shade) curve correpsonds to the EW-CORR switch ``No Photons", red (medium dark shade) curve corresponds to the switch ``CEEX ISR+FSR+EWK" (ISR+FSR here) and blue (dark shade) curve corresponds to the switch ``CEEX FSR+EWK" (FSR only here), as explained in the text.}
\label{fig-2}
\end{figure}

We turn next to the muon pseudo-rapidity ($\eta$) distribution  in Fig.~\ref{fig-2} for the same notational conventions and simulation conditions as we have in Fig.~\ref{fig-1}. 
We see that the effect of the EW corrections on the muon $\eta$ distribution
in the ``FSR only" (CEEX FSR+EWR) and ``ISR+FSR"(CEEX ISR+FSR+EWR) cases are very similar and are significant, so that they must be taken into account in precision LHC physics~\cite{EWsys1,EWsys2,EWsys3,Horace1,Horace2,Horace3}. The inclusive nature of this observable as well as the inclusive nature of the selection cut in this simulation means that we do expect the ``FSR only" and ``ISR+FSR" results to be very close and this is seen here. Similarly, we expect the "No Photons" green (light shade) curve and the black Herwig6.5
curves to agree as they do.\par

One of the important aspects of the MC approach to precision LHC theory is the ability of the MC to give a realistic view, on an event-by-event basis, of the higher order corrections, especially those involving the emission of multiple photon and multiple QCD parton radiation that is observable in the LHC detectors. {\KK}MC-hh now affords this for both the ISR and FSR for the multiple photon radiation in addition to the multiple gluon and quark(anti-quark) radiation in Herwig shower. 
\begin{figure}[ht]
\begin{center}
\includegraphics[width=160mm]{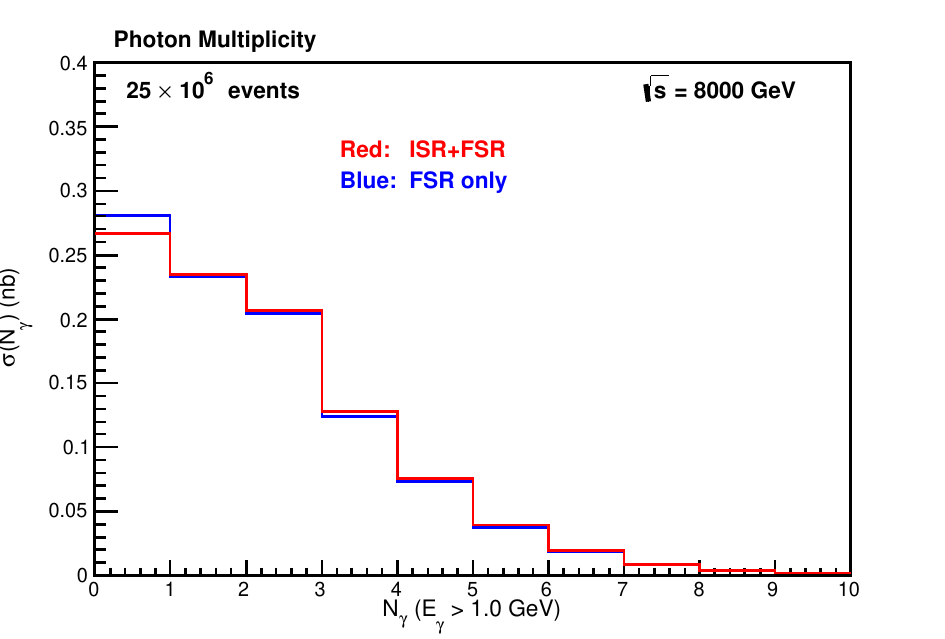}
\end{center}
\vspace{-4mm}
\caption{\baselineskip=11pt The photon number spectrum in {\KK}MC-hh with the cuts specified in the text. The  red (medium dark shade) curve corresponds to the switch ``ISR+FSR+EWK" and blue (dark shade) curve corresponds to the switch ``FSR+EWK", as explained in the text.}
\label{fig-3}
\end{figure}
In Fig.~\ref{fig-3} we show the photon number distribution in the ``FSR only"(``FSR+EWK") and ``ISR+FSR"(``ISR+FSR+EWK") cases as defined above for the same simulation conditions we have in Figs.~\ref{fig-1}-\ref{fig-2}. With the type of statistics that now obtains at the LHC experiments, 
we see that a MC which treats the multiple photon final states in our simulations realistically is essential to determine accurately the responses of the detectors to the EW corrections they do encode. We note that there is additional multiple photon character in the ``ISR+FSR" case compared to the ``FSR only" case in this connection.\par

\begin{figure}[ht]
\begin{center}
\includegraphics[width=160mm]{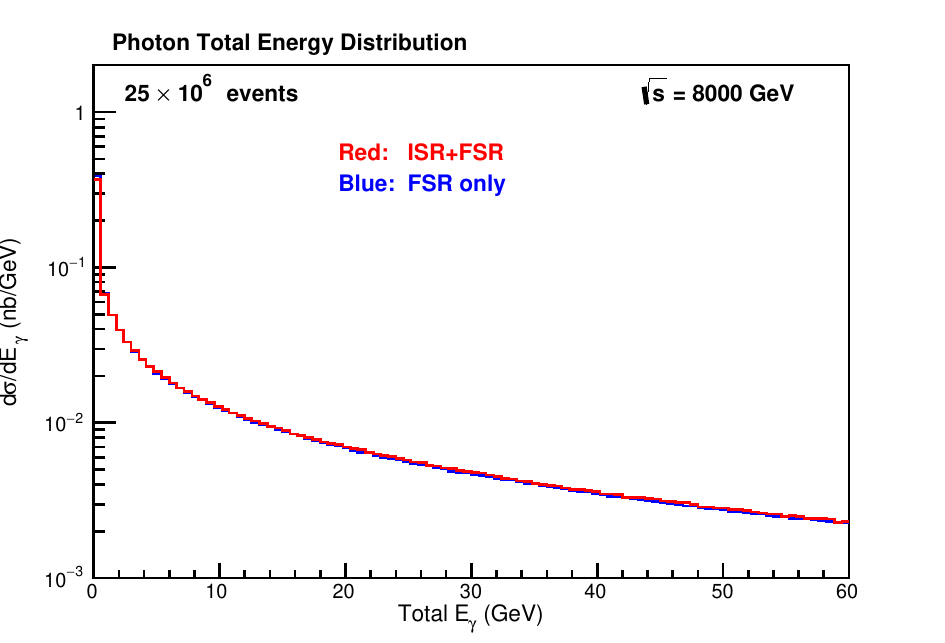}
\end{center}
\vspace{-4mm}
\caption{\baselineskip=11pt The total photon energy spectrum in {\KK}MC-hh with the cuts specified in the text. The  red (medium dark shade) curve corresponds to the switch ``ISR+FSR+EWK"(ISR+FSR) and blue (dark shade) curve corresponds to the switch ``FSR+EWK"(FSR only), as explained in the text.}
\label{fig-4}
\end{figure}

In Fig.~\ref{fig-4} we show the distribution of the total radiated photon energy in our simulations with {\KK}MC-hh with the cut and simulation conditions as described above. 
As expected, the case
corresponding to ``ISR+FSR" shows more photon radiated energy to higher values of energy than does the ``FSR only" case. Again,
for the detector that can resolve photons of the attendant energy, it is essential to have a realistic view of where they actually are in the detector. 
Such a view is afforded by {\KK}MC-hh.\par

We turn next to the total photon $p_T$ spectrum in our {\KK}MC-hh simulations with the cut and conditions as described in Figs.~\ref{fig-1}-\ref{fig-4}. This is a view of the contribution of these photons to the muon pair $p_T$ via recoil: in the EW hard process, the muon pair $\vec{p}_T$ is just the negative of the the total photon $\vec{p}_T$. 
\begin{figure}[ht]
\begin{center}
\includegraphics[width=160mm]{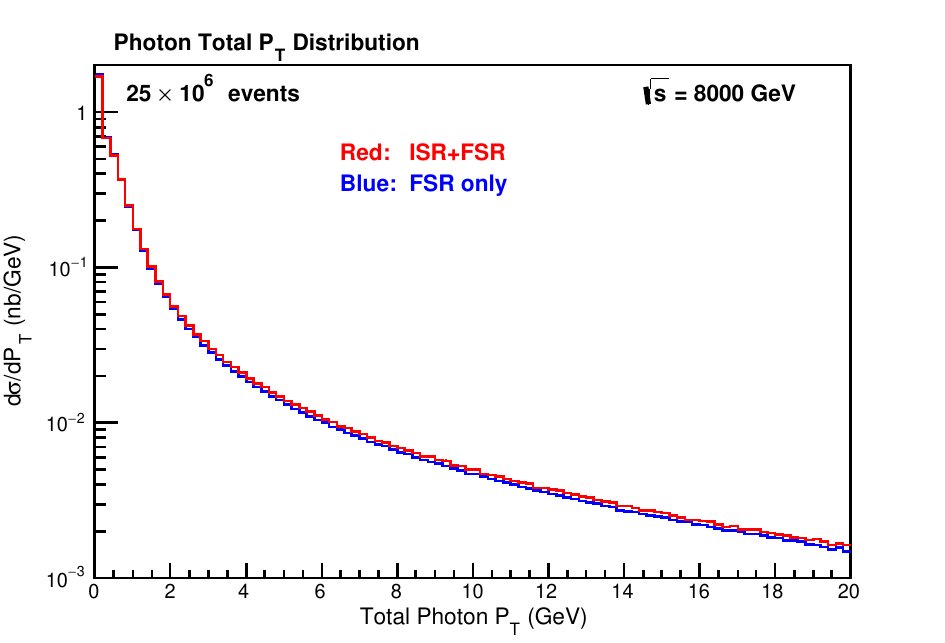}
\end{center}
\vspace{-4mm}
\caption{\baselineskip=11pt The total photon $p_T$ spectrum in {\KK}MC-hh with the cuts specified in the text. The  red (medium dark shade) curve corresponds to the switch ``ISR+FSR+EWK"(ISR+FSR) and blue (dark shade) curve corresponds to the switch ``FSR+EWK"(FSR only), as explained in the text.}
\label{fig-5}
\end{figure}
We see in Fig.~\ref{fig-5}
that most of the spectra lie below 4.0 GeV/c with a larger fraction of events at very small $p_T$ in the ``FSR only" case compared to the ``ISR+FSR" case. For a bin with of 2 GeV/c
for the muon pair $p_T$, our results suggest this larger fraction would be partly compensated so that the difference between the two cases would be reduced accordingly. For less inclusive cuts
than what we have taken here, the situation would need further study~\cite{elswh}.\par

\subsection{Theoretical Comparisons}
The results in Figs.~\ref{fig-1} - \ref{fig-5} give a sample of the type of phenomena that {\KK}MC-hh affords for investigation in the context of having {\em realistic} analysis, especially in the context of detector simulations, for multiple photon radiative effects in $Z/\gamma^*$ production and decay to lepton pairs at LHC and at FCC~\cite{fcc}. There is a considerable
literature on the general area of this subject -- see for example Ref.~\cite{vicini-dorw} for a survey and set of comparisons that were recently completed. Here, we make some contact with  this literature via comparison with the results from the program HORACE~\cite{Horace1,Horace2,Horace3}, where we note that, for example,
in Ref.~\cite{vicini-dorw} one can see how HORACE compares in the general survey of the literature in this latter reference. We will make a more comprehensive 
set of comparisons with the literature elsewhere~\cite{elswh}.\par

In our comparisons with HORACE which follow we turn off the Herwig6.5210  shower. We take the same PDF's and cut for HORACE as we have used above for {\KK}MC-hh: MSTW2008 PDF's and the cut $50 \text{GeV} < M_{q\bar{q}} < 200 \text{GeV}$.
HORACE is run with with exponentiation and its ``best'' EW scheme, using the
same input parameters as {\KK}MC-hh, except that the $Z$ boson mass and width 
are adjusted as in Ref.~\cite{vicini-dorw} to account for the fixed-width 
scheme used in HORACE.
We operate two different levels of precision in {\KK}MC-hh in our tests against HORACE. Specifically,  we employ the best ``CEEX ISR+FSR+EWR"(ISR+FSR) switch
and the ``CEEX FSR+EWK"(FSR only)$\equiv$(FSR) switch as described above.
For HORACE, we then run it with exact \Order{\alpha} with its QED shower for FSR(\Order{\alpha} QED Shower FSR). Thus we would expect in general good agreement between the HORACE results and 
{\KK}MC-hh for the CEEX FSR+EWK (FSR) results, based on the corrections which they entail.
What we find is shown in Tab.~\ref{tab-2}, where we use here the {\KK}MC-hh  CEEX ISR+FSR result as the reference with the same column definitions as we have in Tab.~\ref{tab-1}.
\begin{table}[ht]
\caption{Unshowered tests with HORACE.}
\begin{center}
\begin{tabular}{|c|c|c|c|}
\hline
\text{MC}& \text{EW-CORR}& \text{XSECT}& $\Delta(\text{Rel})$\\
\text{{\KK}MC-hh}& \Order{\alpha^2 L'}+\Ordpr{\alpha^2} \text{CEEX ISR+FSR}  & 993 $\pm$ 1 pb& $\cdots$ \\
\text{{\KK}MC-hh}& \Order{\alpha} \text{CEEX FSR} & 991 $\pm$ 1 pb& (-0.20\%)\\
\text{HORACE}& \Order{\alpha} \text{QED Shower FSR}& 1009.63 $\pm$ 0.40 pb& (+1.67\%)\\
\text{HORACE}& None&1025.22 $\pm$ 0.40 pb&(+3.24\%).\\
\hline
\end{tabular}
\end{center}
\label{tab-2}
\end{table}
We see that, without detailed tuning, the agreement is at the 1.9\% level. We have no reason to believe a fully tuned comparison\footnote{By fully tuned comparison we mean a comparison over a representative set of observables for the equivalent respective input parameter sets and renormalization scheme(s) with the identical phase space constraints, such as what was done in the recently completed analysis in Ref.~\cite{vicini-dorw}.}, which will appear elsewhere~\cite{elswh}, will not show a much closer agreement. We note that the two HORACE results in Table~\ref{tab-2} correspond to simulations with 100 million events while the {\KK}MC-hh results
correspond to simulations with $25\times 10^6$ events.\par

With the understanding that our results that now follow are not yet tuned, so that they should be thought of as guides to where the tuning might focus,
we will now present comparisons of differential spectra with HORACE and the two precision levels in {\KK}MC-hh illustrated in Tab.~\ref{tab-2}. \par

\begin{figure}[H]
\begin{center}
\includegraphics[width=160mm]{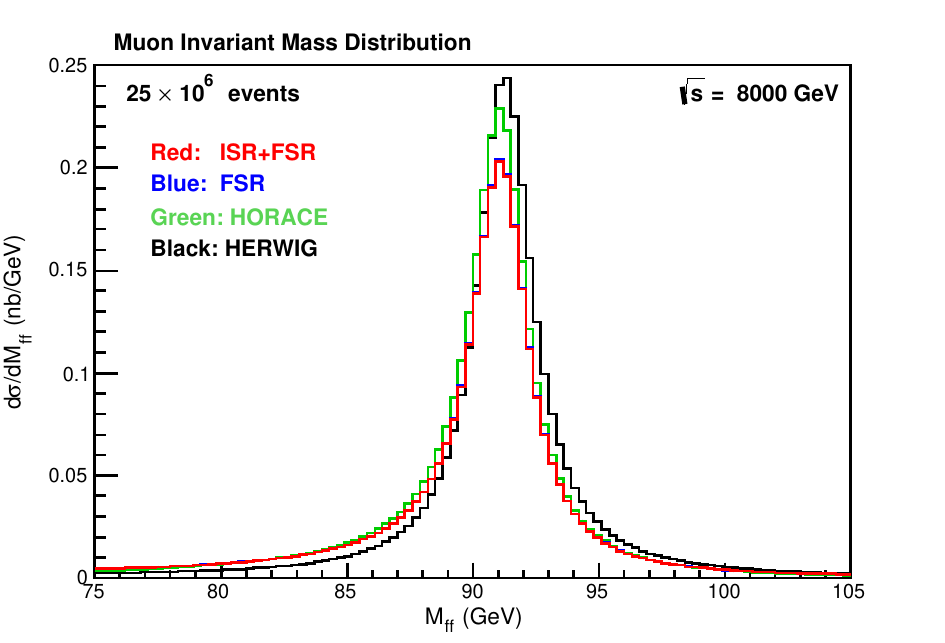}
\end{center}
\vspace{-4mm}
\caption{\baselineskip=11pt The muon pair spectrum in {\KK}MC-hh and HORACE with the cuts specified in the text. The red(medium dark shade) curve corresponds to the EW-CORR switch CEEX ISR+FSR+EWK (ISR+FSR) for {\KK}MC-hh, the blue (dark shade) curve corresponds to the switch CEEX FSR+EWK (FSR)for {\KK}MC-hh and the green(light shade) curve corresponds to the switch \Order{\alpha} QED Shower FSR for HORACE, as explained in the text. For reference, we also show the unshowered Herwig result in the black curve.}
\label{fig-7}
\end{figure}

We start with the muon pair invariant mass distribution, which we show in Fig.~\ref{fig-7} with same simulation conditions and cut as we have Fig.~\ref{fig-1}, but with the shower
turned-off in {\KK}MC-hh and in Herwig6.5, the latter of which is shown in the black curve for reference. All of the simulations have 25 million events.
We see that the two {\KK}MC-hh results are close but that there is some difference with the HORACE result, at the level of $\sim 10\%$ on the peak, for example, between the blue (dark shade) and green (light shade) curves for {\KK}MC-hh  CEEX FSR+EWK and HORACE \Order{\alpha} QED Shower FSR +EWK, respectively, somewhat more than we expect will be the case after some tuning.\par

\begin{figure}[H]
\begin{center}
\includegraphics[width=160mm]{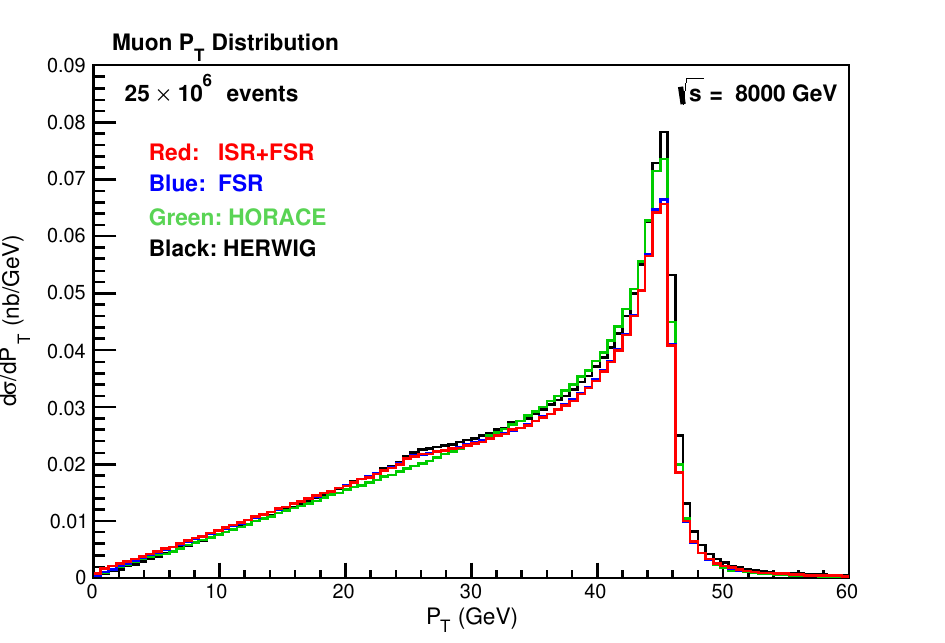}
\end{center}
\vspace{-4mm}
\caption{\baselineskip=11pt The muon $p_T$ spectrum in {\KK}MC-hh and in HORACE with the cuts specified in the text. The labelling conventions and notation are the same as those in Fig.~\ref{fig-7}.}
\label{fig-8}
\end{figure}

Turning next to the comparison for the muon $p_T$ spectrum, we show in Fig.~\ref{fig-8} the results from {\KK}MC-hh, HORACE and HERWIG6.5 with the same simulation conditions, cuts, labeling conventions and notation as in Fig.~\ref{fig-7}. 
We see that there are some differences between the HORACE result and the  CEEX FSR {\KK}MC-hh result. At the peak, we can see a much smaller but nonzero difference between the latter result and the CEEX ISR+FSR {\KK}MC-hh result. This last remark shows that, for the very high precision data at LHC, the most precise {\KK}MC-hh result would 
seem to be preferred.\par

\begin{figure}[H]
\begin{center}
\includegraphics[width=160mm]{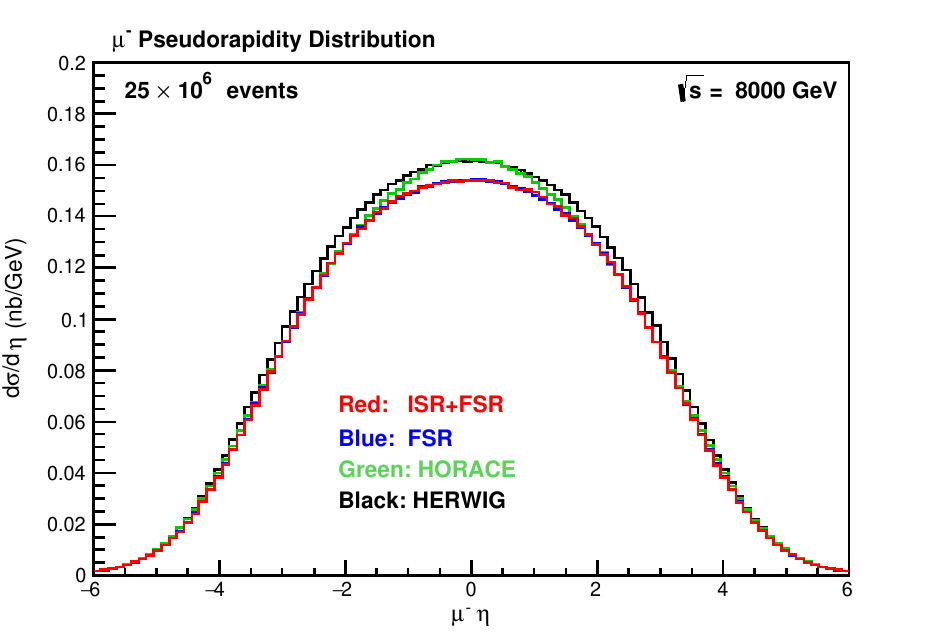}
\end{center}
\vspace{-4mm}
\caption{\baselineskip=11pt The muon $\eta$ spectrum in {\KK}MC-hh and in HORACE with the cuts specified in the text. The labeling conventions and notation are the same as those in Fig.~\ref{fig-7}.}
\label{fig-9}
\end{figure}

For the muon $\eta$ distribution, we show our findings in Fig.~\ref{fig-9}, with the same conditions, conventions and notations as in Fig.~\ref{fig-8} for the two results from {\KK}MC-hh, that from HORACE and that from the reference HERWIG6.5, with all showers turned-off. 
We see again some difference between the HORACE result and the results from the two levels of precision for {\KK}MC-hh, which are very close to each other in this case.\par

\begin{figure}[H]
\begin{center}
\includegraphics[width=160mm]{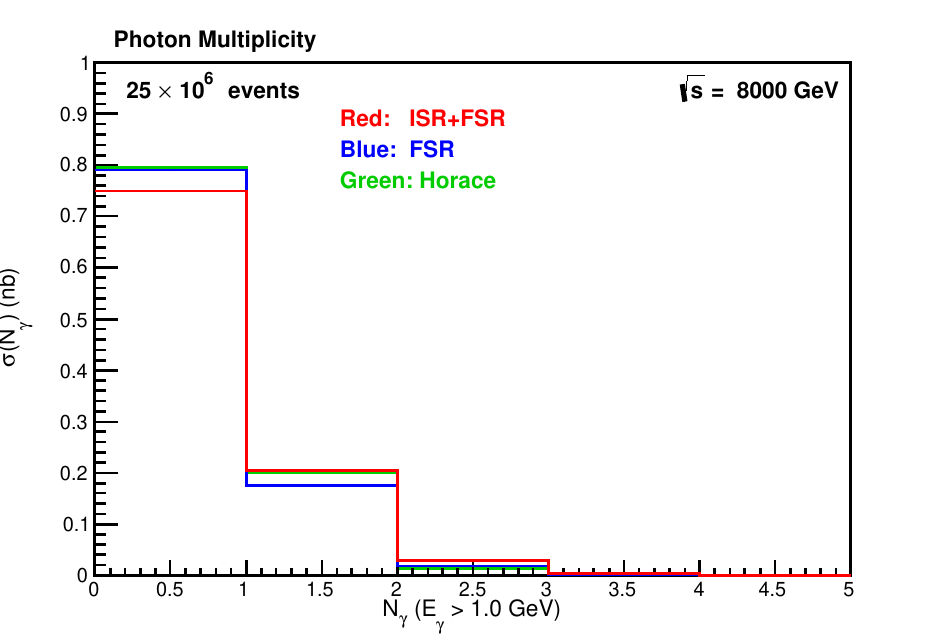}
\end{center}
\vspace{-4mm}
\caption{\baselineskip=11pt The photon multiplicity spectrum in {\KK}MC-hh and in HORACE with the cuts specified in the text. The labeling conventions and notation are the same as those in Fig.~\ref{fig-7}.}
\label{fig-10}
\end{figure}

As we noted above, the multiple photon character of the events interplays with the various efficiencies that may result from detector simulations as needed for precision data analysis for 
large data samples such as those that exist for $Z/\gamma^*$ production and decay to lepton pairs at the LHC. With this latter observation in mind, we take up next in Fig.~\ref{fig-10} the photon multiplicity distribution for the same conditions, conventions and notations as in Fig.~\ref{fig-9} for the two results from {\KK}MC-hh and that from HORACE, with all showers turned-off.  
We see that the distributions for the two FSR only calculations differ significantly from each other and from CEEX ISR+FSR result. \par

\begin{figure}[H]
\begin{center}
\includegraphics[width=160mm]{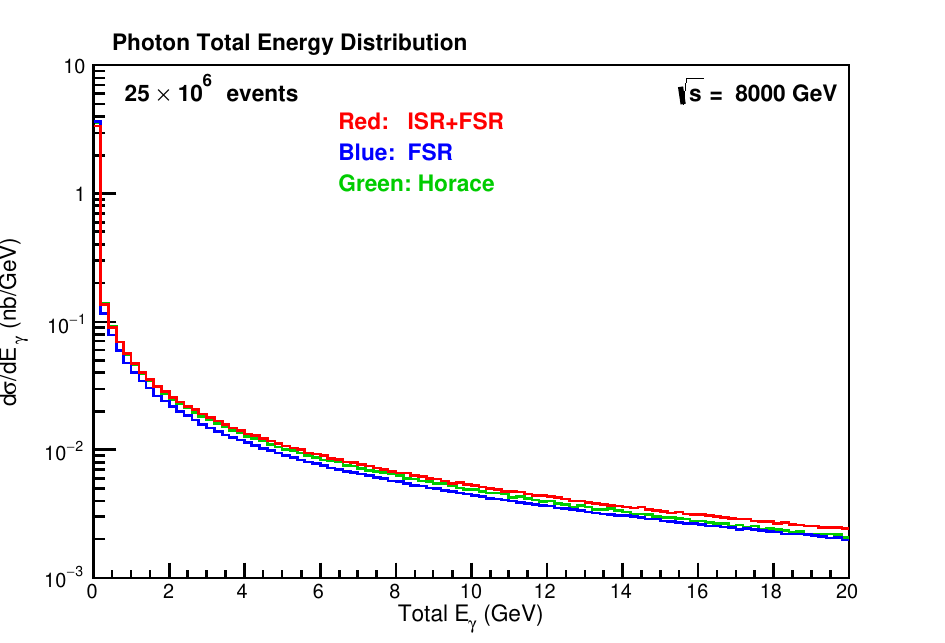}
\end{center}
\vspace{-4mm}
\caption{\baselineskip=11pt The total photon energy spectrum in {\KK}MC-hh and in HORACE with the cuts specified in the text. The labeling conventions and notation are the same as those in Fig.~\ref{fig-10}.}
\label{fig-11}
\end{figure}

With again an eye toward the multiple photon character of the events under study here, we turn next in Fig.~\ref{fig-11} to the total energy radiated into photons for the same conditions, conventions and notations as in Fig.~\ref{fig-10} for the attendant two results from {\KK}MC-hh and that from HORACE. 
We see in this case also that there is some difference between the result for HORACE and the results from the two precision levels of {\KK}MC-hh, which are also different from each other. This situation would seem to suggest that all three of these predictions would give different energy profiles in a precision detector simulation, for example. It would further seem to suggest that the most precise prediction, that for CEEX ISR+FSR, is to be preferred. \par

\begin{figure}[H]
\begin{center}
\includegraphics[width=160mm]{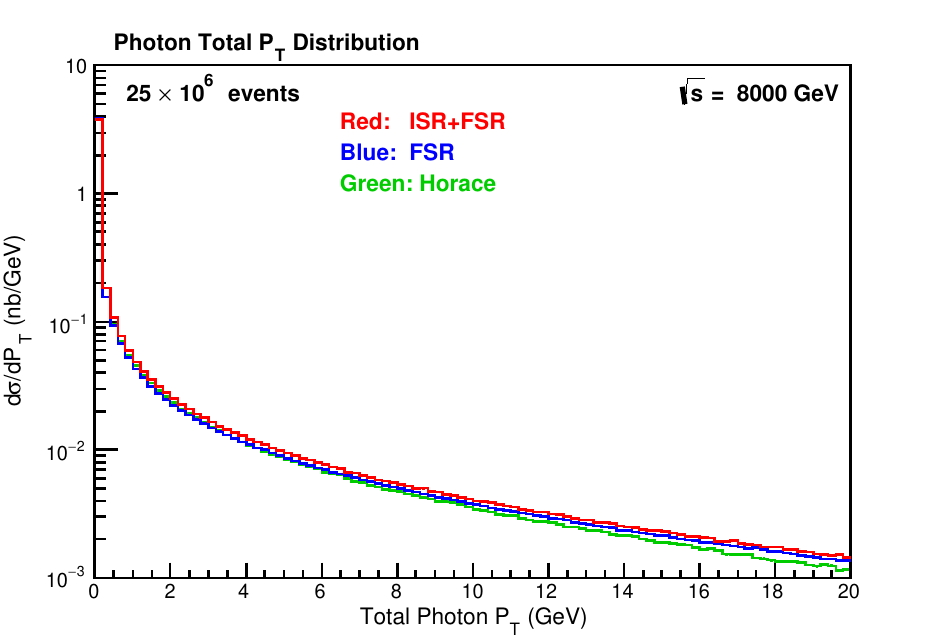}
\end{center}
\vspace{-4mm}
\caption{\baselineskip=11pt The total photon $p_T$ spectrum in {\KK}MC-hh and HORACE with the cuts specified in the text.  The labeling conventions and notation are the same as those in Fig.~\ref{fig-10}.}
\label{fig-12}
\end{figure}

Finally, we recall that the total photon $p_T$ is of interest, as the muon pair recoils against it, so that precision studies would benefit from a precise knowledge of such recoil. We show this spectrum in Fig.~\ref{fig-12} for the same conditions, conventions and notations as in Fig.~\ref{fig-10} for the attendant two results from {\KK}MC-hh and that from HORACE. 
We see that there is some difference between the three predictions and detailed detector simulation could be sensitive thereto, in principle~\cite{elswh}.
Again, the most precise CEEX ISR+FSR would appear to be preferred.\par

\subsection{Consistency Between EEX and CEEX}
We have used the CEEX mode of {\KK}MC-hh in the preceding discussion. As EEX is closer to the QED shower approach to QED resummation that is used in HORACE,
it is important to show the consistency between our EEX and CEEX realizations. 
We now turn to the cross-checks between the best precision EEX and the best precision CEEX, again in the context
of the same conditions as we have Fig.~\ref{fig-1} for example. First, concerning the normalizations, we have the results
\begin{equation}
\begin{split}
\text{CEEX2/EXX2} &= 1.00037\;\; (+0.037\%)\cr
\text{EEX3/EEX2}  &= 0.999975\;\; (-0.0025\%),
\end{split}
\label{eq-crss-chk} 
\end{equation}
where we have denoted by CEEX2 the cross section for the CEEX mode of {\KK}MC-hh with the corrections ${\cal O}(\alpha, \alpha L',\alpha^2 L', \alpha^2 {L'}^2)$ retained in the 
CEEX hard photon residuals~\cite{kkmc419a,kkmc419b,kkmc422} and by EEX2 and EXX3 the \Ordpr{\alpha^n} EEX ISR+FSR cross section results, n=2,3, respectively. 
This shows that {\KK}MC-hh can be used in any of these three modes with confidence in the normalization's consistency.\par

\begin{figure}[H]
\centering
\setlength{\unitlength}{1in}
\begin{picture}(6,2.4)(0,0)
\put(0,0.2){\includegraphics[width=2in]{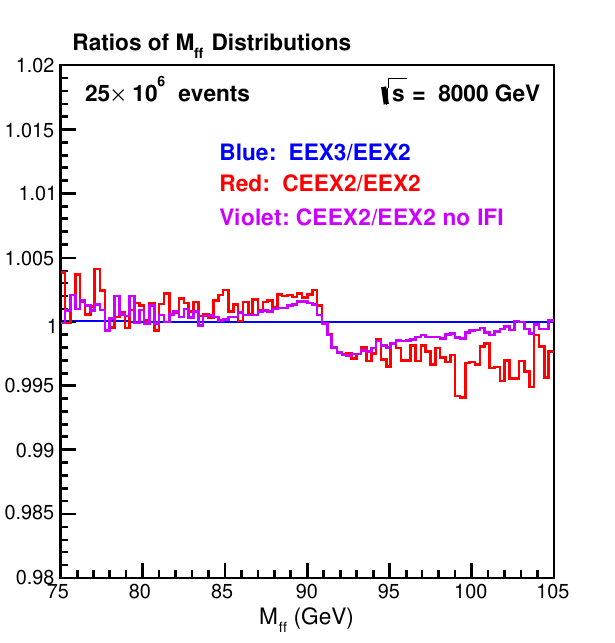}}
\put(2,0.2){\includegraphics[width=2in]{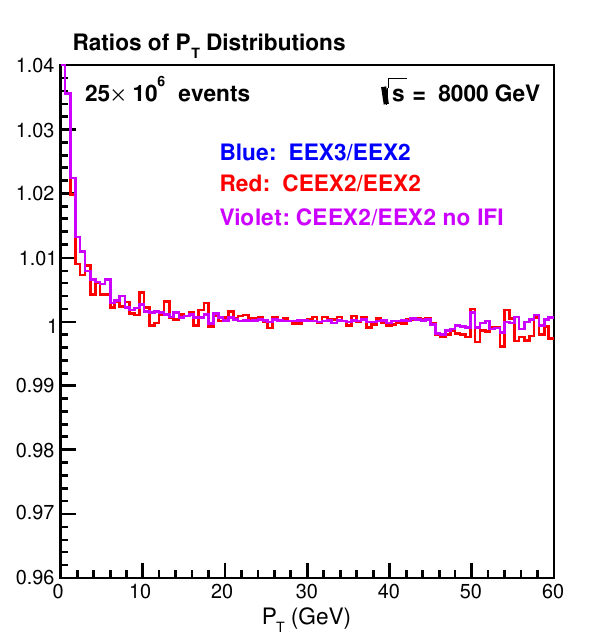}}
\put(4,0.2){\includegraphics[width=2in]{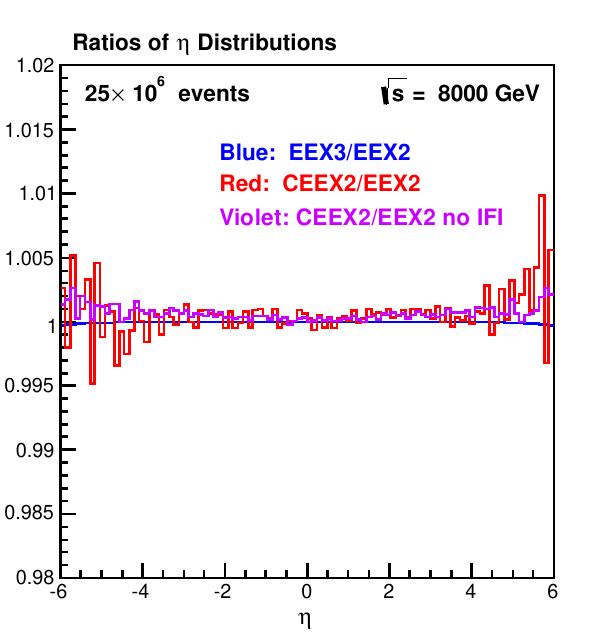}}
\end{picture}
\caption{\baselineskip=11pt Ratios of the muon $p_T$, pair mass, and $\eta$ distributions in {\KK}MC-hh with the cuts specified in the text. The blue (dark shade) curve corresponds to the ratio of EEX3 to EEX2 and the red (medium dark shade) curve corresponds to the ratio of CEEX2 to EEX2. The ratio of CEEX2 without IFI to EEX2 is shown in violet (light dark shade).}
\label{fig-13}
\end{figure}

\begin{figure}[H]
\centering
\setlength{\unitlength}{1in}
\begin{picture}(6,2.4)(0,0)
\put(0,0.2){\includegraphics[width=2in]{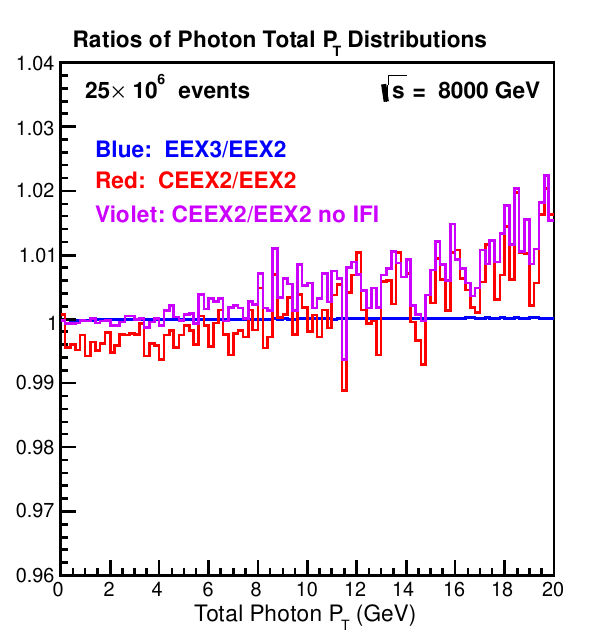}}
\put(2,0.2){\includegraphics[width=2in]{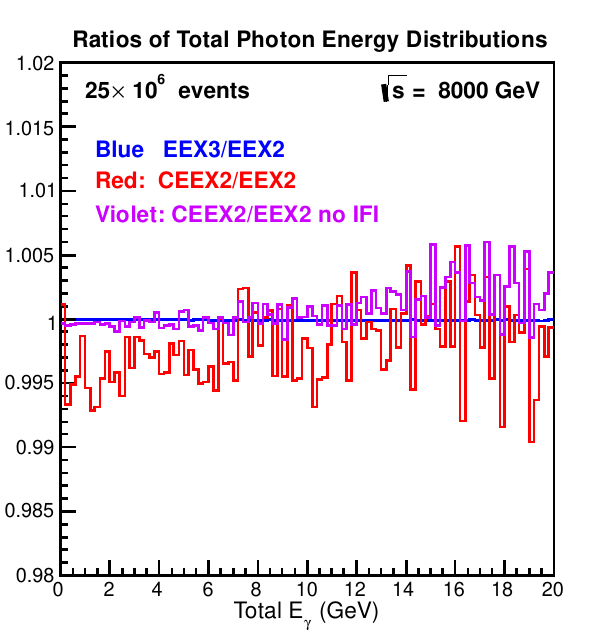}}
\put(4,0.2){\includegraphics[width=2in]{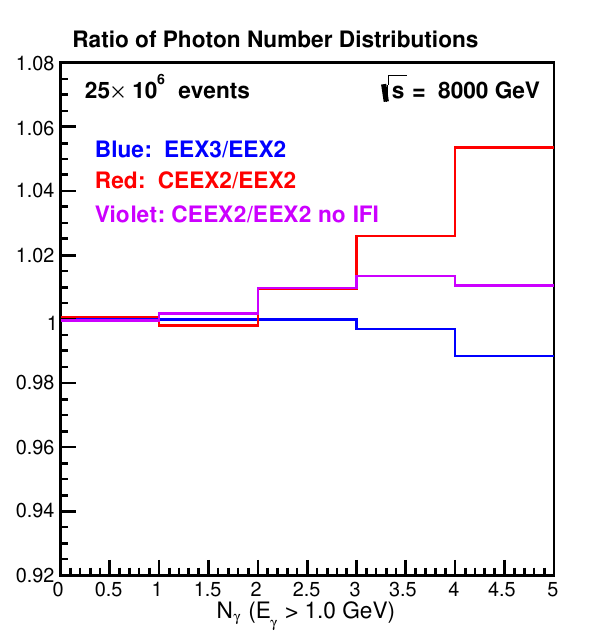}}
\end{picture}
\caption{\baselineskip=11pt Ratios of the photon total $p_T$, energy, and multiplicity distributions in {\KK}MC-hh with the cuts specified in the text. The blue (dark shade) curve corresponds to the ratio of EEX3 to EEX2 and the red (medium dark shade) curve corresponds to the ratio of CEEX2 to EEX2. The ratio of CEEX2 without IFI to EEX2 is shown in violet (light dark shade).}
\label{fig-14}
\end{figure}

Turning now to the analogous distributions which we discussed above, we consider the comparison of the CEEX2 and EEX3 predictions in turn for the muon $p_T$,
muon pair mass, muon $\eta$, photon total $p_T$, photon total energy, and photon multiplicity. We show the muon observables in Fig.~\ref{fig-13} in turn with
the EEX3 prediction in blue (dark shade) and the CEEX2 prediction in red (medium dark shade), where, in view of our results for CEEX2 above, we present the predictions as the
respective ratios of the two predictions to the EEX2 predictions. We also show
the ratio of the CEEX2 prediction with no IFI to the EEX2 prediction in violet (light dark shade) for reference. 
We see very good agreement for these observables.
Similarly, we show in Fig.~\ref{fig-14} the total photon observables in turn with the same format and labeling conventions. 
Again, we see very good agreement for 
these observables, with the clear indication of IFI only in the photon multiplicity distributions for $n > 3$.\par 

\section{Summary}  
 {\KK}MC-hh includes amplitude-based $n\gamma$ emission in single $Z/\gamma^*$ production and decay to lepton pairs for both pp and p\=p colliding beam devices from both the initial and final states in both the EEX and CEEX YFS exponentiation realizations, with the IFI included in the CEEX mode, all in the presence of exact ${\cal O}(\alpha)$ EW corrections from the DIZET library. For the EEX mode, it features \Ordpr{\alpha^3} precision and in the CEEX mode it features 
the sub-leading correction ${\cal O}(\alpha^2 L')$ to \Ordpr{\alpha^2} precision.
The program is still being refined to improve technical matters such as the weight distribution and event generation efficiency. Further comparisons with more of the literature will appear elsewhere. Here, we have made contact with the well-known program HORACE. {\KK}MC-hh is a step toward the goal of an event generator based on nonAbelian QED$\otimes$ QCD resummation and exact ${\cal O}(\alpha_s^2, \alpha_s\alpha, \alpha^2 L')$ hard gluon and hard photon residuals. {\KK}MC-hh is available from the authors upon request.\par
\section*{Acknowledgments}
We acknowledge the hospitality of the CERN Theory Department, which contributed
greatly to the completion of {\KK}MC-hh. S. Yost acknowledges the hospitality
and support of the Theoretical Physics Division of the Institute for Nuclear
Physics of the Polish Academy of Science and a sabbatical funded by The 
Citadel Foundation. S. Yost also acknowledges support from V. Halyo, D. Marlow,
 and Princeton University and the U.S. Department of Energy during the 
development of HERWIRI2.0.\par

\end{document}